\documentclass[preprint,pre,aps,amsmath,amssymb,floatfix]{revtex4}

\usepackage{ulem}
\usepackage{dcolumn}
\usepackage{bm}
\usepackage{color}
\usepackage{graphicx}  
\usepackage{amsmath}
\usepackage{amssymb}

\date{\today}

\begin{document}

\title{\large Twist/Writhe Partitioning in a Coarse-Grained DNA Minicircle Model}

\author{Mehmet Sayar}
\affiliation {Ko\c c University, College of Engineering, Istanbul, Turkey}

\author{Bar{\i}\c s Av\c saro\u glu$^{\dagger *}$}
\affiliation {Ko\c c University, College of Engineering, Istanbul, Turkey}
\altaffiliation{Current address: Martin A. Fisher School of Physics,
  Brandeis University, Waltham, Massachusetts 02453, U.S.A.}

\author{Alkan Kabak\c c{\i}o\u glu$^{\ddagger}$}

\affiliation {Ko\c c University, 
  College of Arts and Sciences, Istanbul, Turkey}

\begin{abstract}
  Here we present a systematic study of supercoil formation in DNA
  minicircles under varying linking number by using molecular dynamics
  simulations of a two-bead coarse-grained model. Our model is
  designed with the purpose of simulating long chains without
  sacrificing the characteristic structural properties of the DNA
  molecule, such as its helicity, backbone directionality and the
  presence of major and minor grooves. The model parameters are
  extracted directly from full-atomistic simulations of DNA oligomers
  via Boltzmann inversion, therefore our results can be interpreted as
  an extrapolation of those simulations to presently inaccessible
  chain lengths and simulation times. Using this model, we measure the
  twist/writhe partitioning in DNA minicircles, in particular its
  dependence on the chain length and excess linking number. We observe
  an asymmetric supercoiling transition consistent with
  experiments. Our results suggest that the fraction of the linking
  number absorbed as twist and writhe is nontrivially dependent on
  chain length and excess linking number. Beyond the supercoiling
  transition, chains of the order of one persistence length carry
  equal amounts of twist and writhe. For longer chains, an increasing
  fraction of the linking number is absorbed by the writhe.
\end{abstract}

\maketitle

\subsection{\label{intro}Introduction}

Conformational features and mechanical properties of DNA {\it in vivo} (such
as supercoil formation, bend/twist rigidity) play an important role in its
packing, gene expression, protein synthesis,~\cite{Luger1997,Halford2004}
protein transport,~\cite{Elf2007} etc. Advances in single-molecule probing and
monitoring techniques in the last decade have provided new opportunities for
detailed analysis of such mechanical and structural properties. The mechanical
response of single DNA molecules,~\cite{Strick2000,Mathew-Fenn2008} the
lifetime of denaturation bubbles,~\cite{Altan-Bonnet2003} and the details of
supercoil formation~\cite{Fogg2006,Forth2008} can now be investigated. These
new experimental findings also triggered renewed theoretical interest in DNA
physics.~\cite{Norouzi2008,Bar2007,Alim2007,Baiesi2009,Rudnick2008,Kabakcioglu2009}

The degree of supercoiling in DNA depends on various tunable parameters, such
as the linking number, external torque and salt concentration. These
parameters essentially modify the molecule's relative preference for twisting
{\it vs} writhing. DNA minicircles provide a convenient setting where
mechanical properties, such as the asymmetry of the molecule under positive
and negative supercoiling, the possibility of torsion-induced denaturation and
kink formation can be studied both experimentally~\cite{Fogg2006} and
theoretically.~\cite{Harris2008,Liverpool2008,Trovato2008} Despite steady
progress in the field in recent decades, a full comprehension of these
phenomena is still ahead of us. Analytical models have been successful in
explaining many qualitative aspects of DNA mechanics, however, they are
typically too simplistic to capture fine details, especially in the short
chain limit. Furthermore, their treatment gets difficult in the presence of
thermal fluctuations and nonlinearities.

Computer simulations provide a wealth of information on short DNA chains.
State of the art full-atomistic simulations are capable of microsecond scale
simulations of short DNA oligomers.~\cite{Perez2007a} Simulations of DNA
oligomers have been successfully used to construct a database~\cite{Dixit2005}
analogous to the crystal structure database. Beyond oligomeric DNA molecules,
full-atomistic simulations have also been performed for DNA
minicircles.~\cite{Harris2008} However, they are short of providing a through
exploration of the conformational space even for minicircles as small as
$100$-$300$ basepairs (bps).

In recent years a number of coarse-grained models have been proposed
to study the conformational and mechanical features of DNA at
intermediate length
scales.~\cite{Drukker2000,Knotts2007,Trovato2008,Sambriski2009a} These
models are typically generated in an ad-hoc manner with tuning
parameters to match some of the key features of DNA, such as the
pitch, persistence length, melting temperature, and sequence
specificity. Drukker{\&}Schatz~\cite{Drukker2000} studied DNA
denaturation of short B-DNA oligomers ($10$-$20$ bps) using a two-bead
model and no major and minor grooves. Knotts {\it et
  al.}~\cite{Knotts2007} introduced a three-bead model to study the
melting dynamics in chains of length $\sim 60$ bps. Their model also
yields a persistence length of $\sim 20$ nm compared to the
experimentally measured value of $\sim 50$ nm. A more recent
model~\cite{Sambriski2009a} where they study DNA renaturation events
is also tuned to reproduce the persistence length. Trovato {\it et
  al.} recently proposed a single-bead model to study thermal melting
that also exhibits major and minor grooves. In simulations of $92$ and $891$
bps long DNA chains, they demonstrated supercoiling and denaturation
using one sample with positive and negative torsional stress for each
length. Here, we present our results on the twist/writhe partitioning using a
novel coarse-grained DNA minicircle model. Using this model we perform
a systematic study of the supercoiling behavior of DNA, and in
particular we investigate the equilibrium amounts of twist and writhe
accommodated by the chain as a function of the applied torsional
stress and the chain length.

Unlike earlier G$\bar{\mbox{o}}$-like approaches, the model parameters
are extracted from full-atomistic DNA simulations via Boltzmann
inversion, with no fitting for structural or mechanical
properties. With only two beads the model captures most structural
details of DNA, such as:
\begin{itemize}
  \item the helicity and the pitch, 
  \item backbone directionality,
  \item major and minor groove structure which results in the anisotropic
    bending rigidity,
  \item persistence length.
\end{itemize}
The accuracy of our coarse-grained model is mostly limited by the accuracy of
the force-field used in the full-atomistic DNA simulations. The
coarse-graining method employed here can be extended in a straightforward
manner to include further details, such as basepair specificity,
hybridization, and explicit charges. However, setting the efficiency of the
model as the priority for mesoscale simulations, we postpone these extensions
to a future study. In the next sections, we outline the model and then
present our results on the twist/write partitioning in DNA minicircles.

\subsection{\label{model}Model}

\begin{figure}
  \begin{center}
      \includegraphics{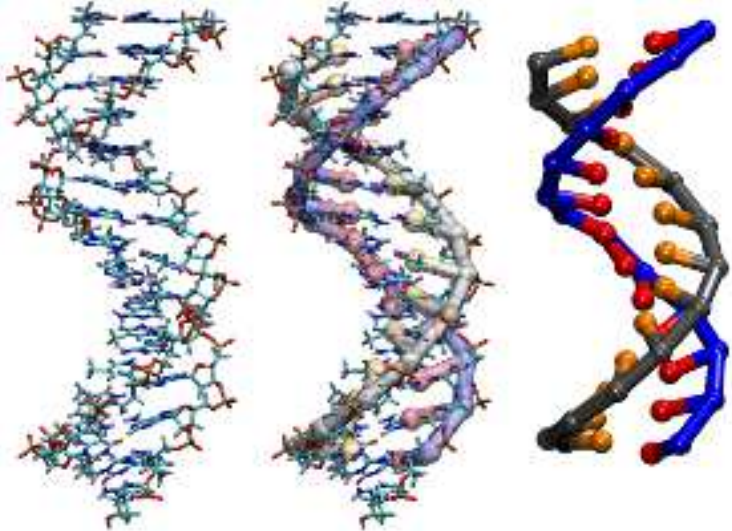}
  \end{center}
  \caption{\label{fatocg} A schematic representation of the mapping
    from the full-atomistic DNA structure (left) to a coarse-grained
    DNA model (right). Note that some of the bonds are omitted
    in the rightmost figure for visual clarity.}
\end{figure}
\subsubsection{Coarse-graining procedure}
The model is composed of two types of ``superatoms'' $P$ and $B$ per
nucleotide, representing the collective motion of the backbone phosphate group
+ sugar (P) and the nucleic acid base (B) as depicted in
Fig.~\ref{fatocg}. Since a realistic description of the helical structure is
sought within a minimalist setting the $B$ superatoms are considered generic,
with no base specificity. The extension to four different bases is
straightforward with an approximately 4-fold increase in the number of model
parameters.  Above simplification is compensated by choosing the superatom
positions optimally, with the criterion that the equilibrium distributions
associated with the degrees of freedom of $B$-superatoms have maximal overlap
when they are calculated for purines ($A$ and $G$) and pyrimidines ($T$ and
$C$) separately. This condition is best met when the Cartesian coordinates of
$P$ superatoms are chosen as the center of mass of the atoms \{ O3', P, O1P,
O2P, O5', C4', O4', C1', C3', C2' \}, while $B$ superatoms are placed at the
center of mass of the atoms \{N9, C8, N7, C5, C6, N3, C4\} for purines and \{
N1, C6, C5, C4, N3, C2\} for pyrimidines.

\begin{figure}
  \includegraphics[width=0.60\textwidth]{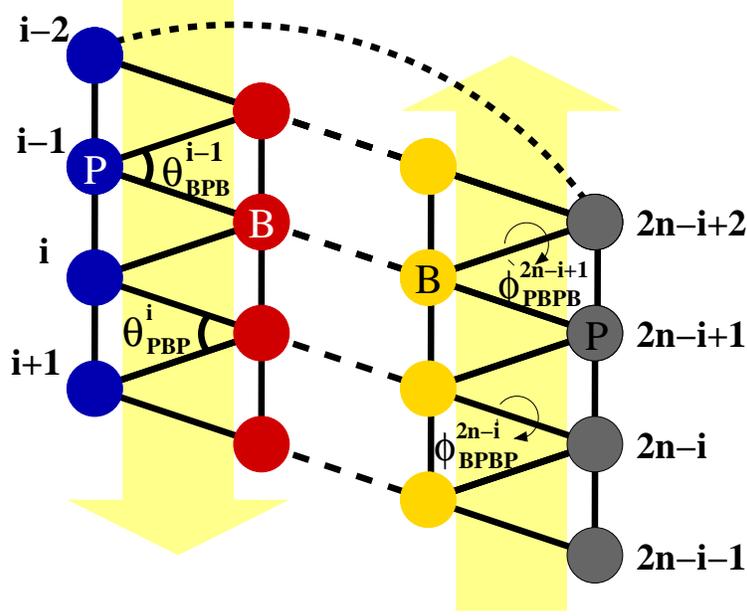}
  \caption{\label{modelfig} Coarse grained model of DNA molecule based on the
  superatoms {P} and {B}, where the former represents the phosphate
  backbone and the sugar group, and the latter represents the
  nucleic-acid base. The superatoms \textbf{$P_i$}, \textbf{$B_i$}
  from the first strand, and the superatoms \textbf{$P_{2n-i}$},
  \textbf{$B_{2n-i}$} from the second strand form the nucleic-acid
  basepair which are connected by hydrogen bonds in the original
  system. The intra-strand bonds \textbf{$P_iB_i$, $B_iP_{i+1}$,
    $P_iP_{i+1}$, $B_iB_{i+1}$} are shown by solid lines. The
  inter-strand bonds $B_iB_{2n-i}$ and $P_iP_{2n-i}$ are shown by
  dashed lines. $\phi_{PBPB}^{i}$ and $\phi_{BPBP}^{i}$ are the
  dihedral angles defined by $P_i$, $B_i$, $P_{i+1}$, $B_{i+1}$ and
  $B_i$, $P_{i+1}$, $B_{i+1}$, $P_{i+2}$, respectively. Similarly,
  $\theta_{BPB}^i$ and $\theta_{PBP}^i$ represent the bond angles
  defined by $B_{i-1}$, $P_i$, $B_i$ and $P_i$, $B_i$, $P_{i+1}$. The
  dihedral angle stiffness is explicitly included in the
  coarse-grained potential, whereas the bond angle stiffness
  is mostly due to the intra-strand $PP$ and $BB$ bonds. }
\end{figure}
Fig.~\ref{modelfig} shows the coarse-grained model composed of the
superatoms {P} and {B} and the adopted indexing convention for the two
strands.

\subsubsection{Interactions}

The effective interactions incorporated into the model are shown in
Fig.~\ref{modelfig}. We use four bonded and two dihedral potentials that maintain the local single-strand geometry:
\begin{itemize}
\item{harmonic bonds $P_iB_i$, $B_iP_{i+1}$, $P_iP_{i+1}$ and
    $B_{i-1}B_{i}$ that fix the intra-strand superatom distances as well as the angles $\theta^i_{PBP}$
    and $\theta^i_{BPB}$,}
\item{dihedral potentials associated with the angles
    $\phi^i_{PBPB}$ and $\phi^i_{BPBP}$.}
\end{itemize}
All four harmonic bond potentials have the form
\begin{eqnarray}
  \label{harmonic_potential}
  V_b(r)&=&\frac{1}{2}K_b{(r-r_0)}^2\ ,
\end{eqnarray}
where the stiffnesses, $K_b$, and the equilibrium bond lengths, $r_0$,
differ as listed in Table \ref{forceconstants}. In particular, the
difference between $P_iB_i$ and $B_iP_{i+1}$ bond parameters reflects
5'-3' directionality of the molecule. The choice of harmonic bond
potentials $P_iP_{i+1}$ and $B_{i-1}B_{i}$ over true angular
potentials increases computational efficiency without significantly
distorting the equilibrium distributions, as verified in Fig.~\ref{distributions} for the BPB bond angle.

The torsional stiffness of the dihedral angles $\phi^i_{PBPB}$ and
$\phi^i_{BPBP}$ defined, respectively, by the superatoms
$P_iB_iP_{i+1}B_{i+1}$ and $B_{i-1}P_{i}B_{i}P_{i+1}$ is modeled by
the potential,
\begin{eqnarray}
\label{ang_potential}
  V_d(\phi)=K_d[1-\cos(\phi-\phi_0)]\ ,
\end{eqnarray}
where, again, the two stiffness coefficients, $K_d$, for $BPBP$ and
$PBPB$ dihedral angles and their equilibrium values, $\phi_0$, are
separately determined from the full-atomistic simulation data as
described below.

In addition to the intra-strand interactions above, we define two
inter-strand potentials that stabilize the double-stranded
structure. First is a tabulated potential connecting the pairs
$B_iB_{2n-i}$ and reflects the hydrogen bonding between the
nucleic-acid bases A-T and G-C (again, base specificity is omitted at
this stage). The use of a tabulated potential may facilitate
monitoring base-pair breaking events in future studies. The second
inter-strand interaction takes into account the steric hinderence of
the base atoms and the electrostatic repulsion of the phosphate groups
through a repulsive potential between the superatoms $P_i$ and
$P_{2n-i}$. This interaction also helps maintain the directional
nature of the hydrogen bonding between the base pairs, which in our
model is displayed by the alignment of the superatoms $P_i$, $B_i$,
$B_{2n-i}$, and $P_{2n-i}$. Functional forms of both interactions are
discussed in the next section.

Finally, a Lennard-Jones excluded-volume potential between all
superatom pairs (except $P_iP_{i+2}$) that do not interact via
previously defined potentials maintains the self-avoidance of the DNA
chain. 

\subsubsection{Determination of the force constants}

The force constants and the equilibrium values for bond and dihedral
potentials are obtained from the thermal fluctuations of the
associated superatoms via Boltzmann inversion.~\cite{Reith2003} The
fluctuation data is obtained from molecular dynamics (MD) trajectories
of a full-atomistic study by Dixit \textit{et al.}~\cite{Dixit2005} The
full-atom MD data includes room-temperature simulations of all
possible tetramers of the nucleic-acid bases located at the center of
a $15$ bps long B-DNA oligomers. Since our model does not include
base-pair specificity, all tetramer data was given equal weight
throughout our analysis.

\begin{figure*}
  \begin{center}
    \begin{minipage}[t]{1.0\textwidth}
      \includegraphics[width=1.0\textwidth]{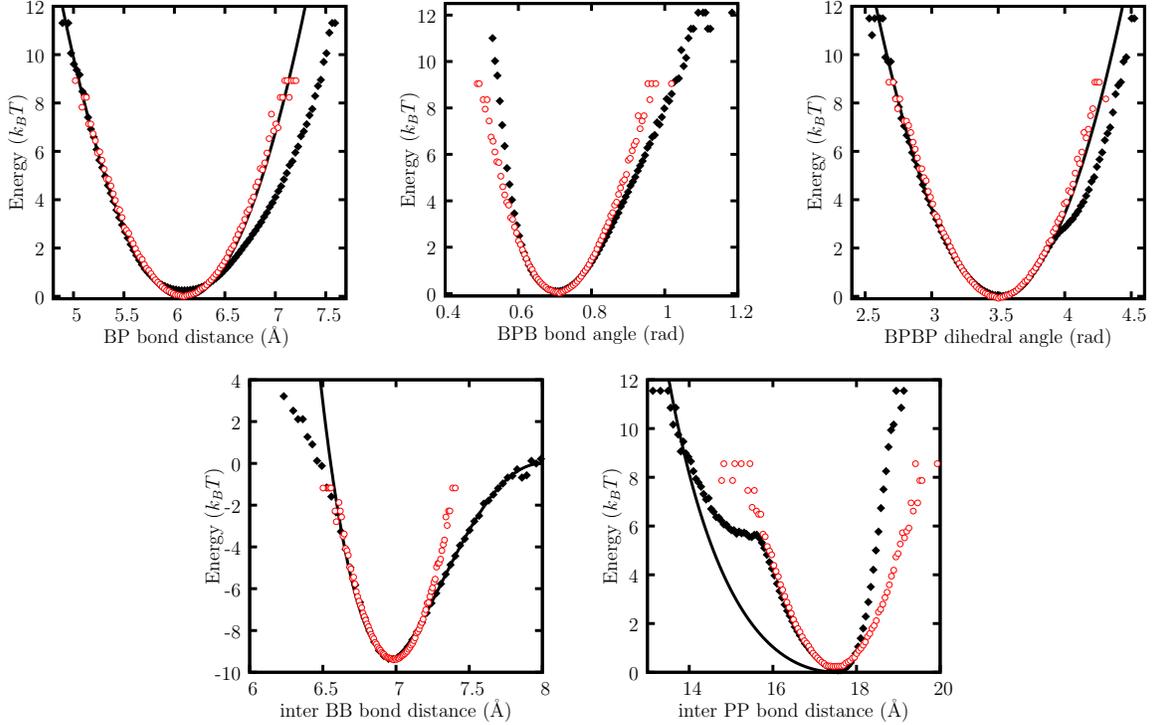}
    \end{minipage}
  \end{center}
  \caption{\label{distributions} Potential of mean force curves for
    intra- and inter-strand interactions obtained by Boltzmann
    inversion of the relevant distance/angle probability
    distributions. The target distributions obtained from the
    full-atomistic trajectories (filled black diamonds), the
    coarse-grained model potentials derived from these (solid black
    lines), and the results of the coarse-grained model simulations
    (open red circles) for the intra- (top) and inter-strand (bottom)
    degrees of freedom. {PB} bond, {PBP} angle, and {PBPB} dihedral
    angle distributions (not shown) yield similar results.}

\end{figure*}

Boltzmann inversion of the probability distributions for each type of
bond length, bond angle, and dihedral angle (obtained from the thermal
fluctuations of superatom centers) yields potentials of mean force
(PMF) which are shown with solid diamond symbols in Fig.~\ref{distributions}.

These PMF curves are used to obtain the force constants for the
intra-strand bonded interactions by means of harmonic fits. The fits
obtained for $r_{BP}$, and $\phi_{BPBP}$ are shown with solid-black
lines in Fig.~\ref{distributions} and the force constants obtained
by these fits are listed in Table \ref{forceconstants}. The anharmonic
features in PMF curves can be captured using more sophisticated
potentials, which could be implemented if a more specific model is
desired. 

\begin{table}[h]
  \begin{tabular}[c]{|l||l|l|l|l|l|l|}
    \hline
    Interaction Type & Equilibrium Position  & Force Constant\\
    \hline
    $P_iB_i$ bond & $r_0$=5.45 \AA & $K_b$=7.04 $k_BT/{\AA }^2$\\
    \hline
    $B_iP_{i+1}$ bond & $r_0$=6.09 \AA & $K_b$=16.14  $k_BT/{\AA }^2$\\
    \hline
    $P_iP_{i+1}$ bond & $r_0$=6.14 \AA & $K_b$=20.36  $k_BT/{\AA }^2$\\
    \hline
    $B_iB_{i+1}$ bond & $r_0$=4.07 \AA & $K_b$=15.93  $k_BT/{\AA }^2$\\
    \hline
    $PBPB$ dihedral & ${\phi}_0$=3.62 rad & $K_d$=25.40  $k_BT/{rad}^2$\\
    \hline
    $BPBP$ dihedral & ${\phi}_0$=3.51 rad & $K_d$=27.84  $k_BT/{rad}^2$\\
    \hline
  \end{tabular}
  \caption{Force constants obtained by fitting the analytical forms in
    Eqs.(\ref{harmonic_potential}) and (\ref{ang_potential}) to the Boltzmann inverted distributions shown in Fig~\ref{distributions}. }
  \label{forceconstants}
\end{table}

Also shown in Fig.~\ref{distributions} are the PMF curves for the
inter-strand bond distances between $B_iB_{2n-i}$ and $P_iP_{2n-i}$, where a
strong asymmetry is evident in both. The interaction among $B_iB_{2n-i}$
superatoms stems from the hydrogen bonds as discussed above, and displays an
equilibrium separation. The $B_iB_{2n-i}$ interactions is incorporated into
the model via a tabulated potential, which is obtained by a smooth curve
fit to the PMF data. 

The $P_iP_{2n-i}$ PMF curve also displays an equilibrium separation. Unlike
the $B_iB_{2n-i}$ interaction, here we only model the repulsive part of this
interaction via a tabulated potential. The attractive part is already captured
by the previously discussed potentials, as we will discuss below.

The excluded volume of the superatoms is represented via repulsive
Lennard-Jones interactions,
\begin{equation}
  \label{LJ}
  U_{LJ}(r)= 
  \begin{cases}
    4 \big[ \big(\frac {r_o} {r} \big)^{12} - \big(\frac {r_o} {r} \big)^{6} + 0.25\big] & r< r_{cut} \\
    0 & r \ge r_{cut} 
  \end{cases}
\end{equation}
measured in units of $k_BT$. For all B-B and B-P pairs $r_o = 5.35
\, \AA $ and $r_{cut}=6 \, \AA $, whereas for all P-P pairs these
values are doubled. Superatom pairs that are bonded and all
$P_iP_{i+2}$ pairs are excluded from these Lennard-Jones
interactions. The constants and exclusions for these Lennard-Jones
interactions are chosen such that intra- and inter-strand interactions
previously defined are not influenced when the Lennard-Jones
interactions are switched on.

The coarse-grained model potential does not include any explicit
electrostatic interactions, therefore strictly speaking this model is
suitable for high salt concentrations.

\subsubsection{MD simulation}

ESPResSo package~\cite{Limbach2006} was used for all coarse-grained MD
simulations with the Langevin thermostat at room temperature. All
super-atoms were assumed to have the same mass roughly equivalent to 170
atomic mass units. The length and energy units were set to 1 \AA ~and
$k_BT$ ($T$=$300$ K), respectively, from which the unit time ($\tau$)
can be determined by dimensional analysis to be approximately 7 fs. The
equations of motion were integrated by using the velocity Verlet
algorithm with a time-step of 0.015 $\tau$.

We simulated both linear and circular DNA chains. The initial
configurations were chosen as their respective ground-states obtained
through over damped MD simulations. The data collection was performed
after thermal equilibration, where the required equilibration time
depended on the measured quantity. For example, the equilibrium
distributions for the degrees of freedom of the coarse-grained model
(Fig.~\ref{distributions}) took approximately 10 CPU minutes on an
Intel Quad-Core machine using a single processor, whereas thermal
averages of the writhe and the twist (Fig.~\ref{writhe}) for the
longest circular DNA we considered required upto 64 CPU weeks using 8
processors in parallel. MD trajectories are visualized by VMD
package.~\cite{Humphrey1996}

\subsubsection{Equilibrium Properties}
\begin{figure}
  \includegraphics{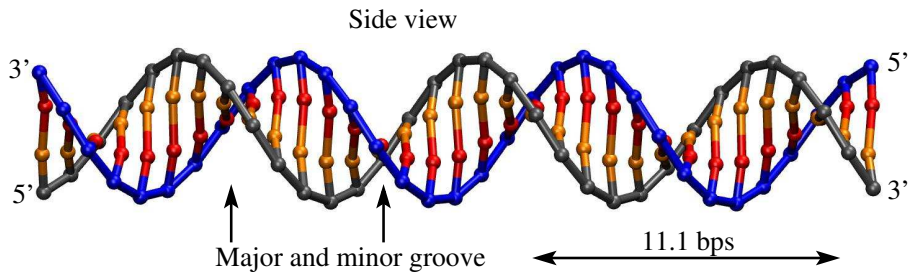}
  \caption{\label{groundstate}Ground state structure of the model DNA.
    and top (right) view of the equilibrium structure.  Key structural
    features of the DNA molecule, such as the directionality of the
    backbone and major and minor groves are present in the coarse grained
    model. Note that some of the bonds are omitted in the figure
    for visual clarity.}
\end{figure}

The ground-state structure of the molecule is shown in
Fig.~\ref{groundstate}. Despite the fact that only two beads are used
for each unit, the model successfully captures most of the essential
features of the DNA molecule.  The directionality of the strands is
reflected in the differences between the force constants for BP
\textit{vs.} PB bonds and PBPB \textit{vs.} BPBP dihedral angles. The
double helical structure with major and minor grooves is also
captured.  The helical pitch (the method of calculation is discussed
further below) is 11.4 bps in ground state and drops to 11.1 bps at
room temperature, suggesting an anharmonic twist rigidity which is a
complex function of the model potentials given above. The reason for
the somewhat higher helical pitch we find here in comparison with the
actual DNA is discussed at the end of this section.

We tested the derived coarse-grained potentials by performing an MD
simulation of a 36 bps long linear chain. The resulting Boltzmann
inverted distributions of the bond lengths, bond angles and dihedral
angles are shown with open red circles in
Fig.~\ref{distributions}. The distributions associated with
intra-strand interactions nicely match the potentials used, which
themselves are the best harmonic fits to the corresponding
full-atomistic data.  The fact that the given potentials are recovered
from the thermal fluctuations of the superatoms suggests that the
degrees of freedom used in the coarse-grained Hamiltonian are
minimally coupled. Also shown in Fig.~\ref{distributions} is the
inverted form of the BPB bond angle distribution (top row, in the
middle). The agreement between the full-atomistic (filled black
diamonds) and the coarse-grained (open red symbols) simulation
results is observed also for other angular potentials (not shown) and
justifies the aforementioned use of harmonic bond potentials as a
replacement for true angular potentials.

On the other hand, the inter-strand degrees of freedom, namely
$B_iB_{2n-i}$ and $P_iP_{2n-i}$, display some degree of coupling to
other bonded interactions. $B_iB_{2n-i}$ interaction represents the
hydrogen bonding within a base-pair. The tabulated potential we used
faithfully reproduces the equilibrium fluctuations upto a few $k_BT$
(see bottom-left graph in Fig.~\ref{distributions}). Higher-energy
excitations of the $B_iB_{2n-i}$ bond are suppressed beyond the level
imposed by the derived potential, suggesting a coupling with the other
degrees of freedom in this regime.

$P_iP_{2n-i}$ interaction we have used is a purely repulsive
interaction, as described above. The associated tabulated potential
had to be iteratively softened until a good match was obtained with
the full-atomistic data on the left of the Boltzmann inverted
distributions (for $r\lesssim 17.5\,\AA$ in the bottom-right graph in
Fig.~\ref{distributions}). The attractive right-hand-side of
the effective $P_iP_{2n-i}$ potential derived from the superatom
fluctuations is solely due to the remaining interactions.

\begin{figure}
  \includegraphics[width=1.0\textwidth]{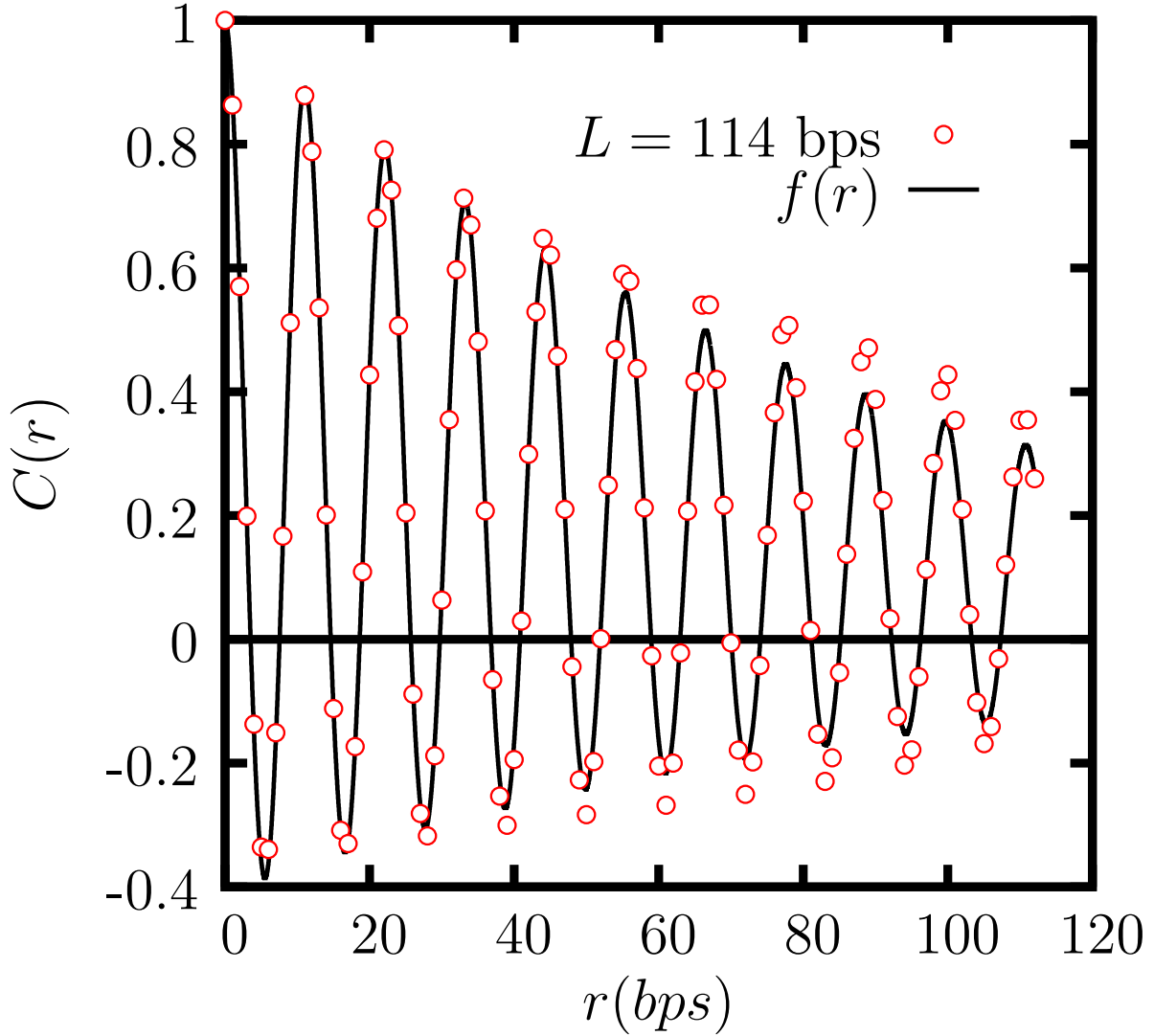}
  \caption{\label{persistence} The correlation of the
    \textbf{$\vec{PP}$} vectors as a function of the vector length
    $r$ and functional fit to obtain the persistence length and pitch of the
    model molecule. }
\end{figure}

We have looked at the persistence length of our model DNA by means of
extended simulations of linear molecules with freely fluctuating ends
and lengths 114, 228, and 456 bps. Snapshots from these simulations
were used to calculate the persistence length and the pitch as
explained below.

Let $\vec{q_i}$ be the vector joining $P_i$ and $P_{i+1}$ and $C(r) =
\langle \vec{q_i}\cdot\vec{q}_{i+r}\rangle$. Fig.~\ref{persistence}
shows $C(r)$ for a $114$ bps long DNA, calculated by a running average
along the chain which is further averaged over 200 snapshots. The
helicity of the molecule results in a sinosoidal function with an
exponentially decaying amplitude using which both the persistence
length and the pitch can be measured. The solid line in
Fig.~\ref{persistence} represents the fit to $C(r)$ via the following
empirical function,
\begin{eqnarray}
  \label{perseqn}
  f(r) = e^{-r/l_p}\,\big[a + (1-a)\cos(2\pi r/\lambda)\big]\ \ ,
\end{eqnarray}
where $a$ is a geometrical constant related to the aspect ratio of the
helix and given by $a= (1/2) - 1/[1+(\lambda/2\pi R)^2]$, $\lambda$ is
the helical pitch and $R$ is the radius measured between the helix
center to the $P$ superatoms. The decay rate of the correlations
measured by fitting Eq.~(\ref{perseqn}) in the interval $0\le r \le
30$ gives the persistence length ($l_p$) of the model DNA as 96 bps
consistently for chain lengths of $114$, $228$, $456$ bps (data not
shown for $228$ and $456$ bps long chains). At larger chemical lengths
($r>30$), finite size effects (short chains) and long equilibration
times (long chains) limit the accuracy of the data.

The persistence length of our model DNA is shorter
that the experimentally measured length of 50 nm's ($\approx 150$
bps). The stiffness of a DNA chain is a result of both the
bonding/angular interactions that form the local helical structure and
the electrostatic self-repulsion imposed by the high line charge
density. The electrostatic interactions are included here only
implicitly, through the coarse-grained force-field parameters that best
fit the full-atomistic simulations performed with explicit
electrostatic interactions. This approximation is one possible reason
for the shorter persistence length we find, while the imperfection (below) of
the full-atomistic force-field used by Dixit {\it et
  al.}~\cite{Dixit2005} is another. 

Equation (\ref{perseqn}) simultaneously provides an estimate for the
helical pitch at room temperature as $\lambda = 11.1$ bps.  This value
is also slightly higher than that measured ($\simeq 10.5$ bps) for the
B-DNA. An inspection of the full-atomistic data reveals that the same
mismatch exists for the oligomers simulated by the AMBER parm94
force-field~\cite{Cornell1995} used by Dixit and
coworkers.~\cite{Dixit2005} Therefore, the higher pitch is, in fact, a
consequence of the imperfect AMBER parm94 force-field rather than a
shortcoming of the inverted Boltzmann method used here.

\subsection{Twist {\it vs} Writhe in DNA minicircles}

Our main goal in this study is to analyse the twist/writhe partitioning in DNA
minicircles. In addition to serving as a demonstration of the model's
capabilities, this problem is also relevant to the denaturation behavior of
DNA chains under conserved linking
number.~\cite{Rudnick2008,Kabakcioglu2009} In this section, we will explain how
we measure the response of a circular DNA chain to applied torsional stress
and present our results for chains of different lengths under varying stress
levels.
\subsubsection{Calculating twist and writhe on a discrete chain}
Twist and writhe reflect two geometrically distinct modes
of response of a DNA chain to applied torsional stress. Let
$\vec{r}_1(s)$ and $\vec{r}_2(s)$ be the two closed curves
interpolating $B$-superatoms of each strand, parametrized by the
continuous variable $s$ and obtained here by the cubic spline
method. The centerline of the DNA is given by $\vec{r}(s) =
[\vec{r}_1(s) + \vec{r}_2(s)]/2$. The unit tangent and normal vectors
at any point $s$ are
  \begin{eqnarray}
    \vec{t}(s) &=& d\vec{r}(s)/ds \nonumber \\
    \vec{u}(s) &=& (\vec{r}_1(s)-\vec{r}_2(s))/|\vec{r}_1(s)-\vec{r}_2(s)| \ .
  \end{eqnarray}
  Then, the twist ($T\!w$) of the chain which is a measure of the
  sum of the successive basepair stacking angles is formally given
  by:~\cite{Kamien2002}
  \begin{eqnarray}
    \label{twist_eq}
    T\!w &=& \frac{1}{2\pi}\,\oint ds\ 
    \vec{t}(s)\cdot\bigg[\vec{u}(s)\times\frac{d\vec{u}(s)}{ds}\bigg]\ .
  \end{eqnarray}
  The writhe ($W\!r$) is a nonlocal property associated with the
  torsional stress stored in the conformation of the centerline (as in
  the coiling of the old telephone chords) and can be obtained using
  \begin{eqnarray}
    \label{writhe_eq}
    W\!r &=& \frac{1}{4\pi}\,\oint ds \oint ds'\ 
    \vec{t}(s)\times\vec{t}(s')\cdot\frac{\vec{r}(s)-\vec{r}(s')}{|\vec{r}(s)-\vec{r}(s')|^3}\ .
  \end{eqnarray}
  For a DNA chain constrained to have a fixed ``linking
  number''($Lk$), the number of times one chain loops around the
  other, total twist and total writhe are connected by the
  relation~\cite{white1969,fuller1971}
  \begin{eqnarray}
    \label{Fuller_eq}
    Lk &=& T\!w + W\!r\ .
  \end{eqnarray}
  
\begin{figure*} [ht]
  \begin{center}
      \includegraphics[width=1.10\textwidth]{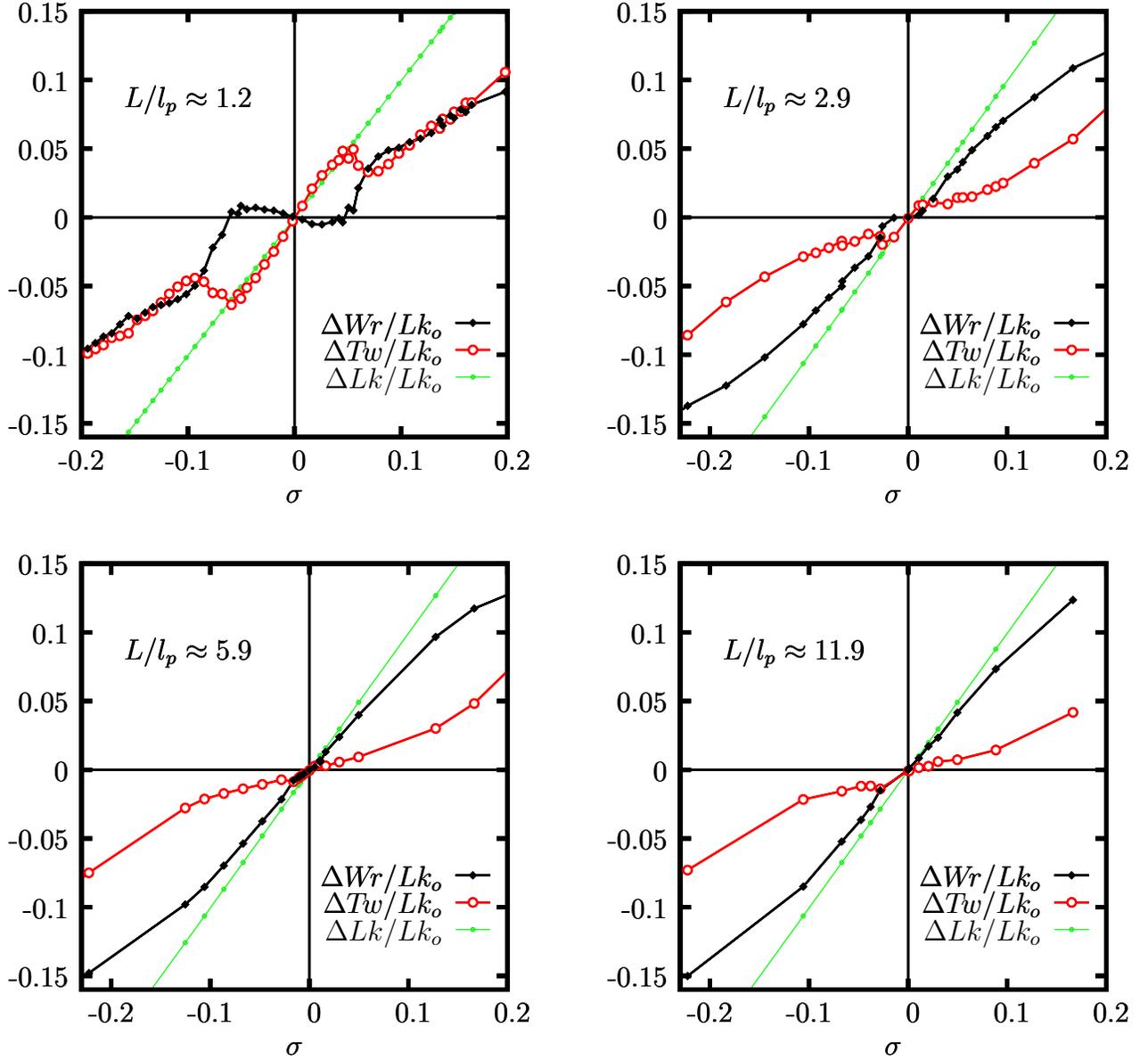}
  \end{center}
  \caption{\label{writhe} The ensemble averaged writhe (black) and
    twist (red) densities of circular DNA molecules for different
    excess linking numbers $\sigma=(Lk/Lk_0)-1$. Also shown in green
    is the sum of the two, confirming the agreement with
    Eq.~(\ref{Fuller_eq}). Each graph corresponds to a different
    minicircle size out of $L = 1.2$, $2.9$, $5.8$, and $11.9$
    $\times$ (persistence length).}
\end{figure*}

  A variety of methods have been proposed for calculating the amount of $T\!w$
  and $W\!r$ on discrete chains.~\cite{Klenin2000,deVries2005} We found that,
  constructing the curves $\vec{r}_{1,2}(s)$ by using cubic splines and
  numerically evaluating Eq.~(\ref{twist_eq}) and Eq.~(\ref{writhe_eq}) is the
  most accurate approach that guarantees validity of Eq.~(\ref{Fuller_eq}) at
  all times. Each snapshot that was used for calculating the average values of
  $T\!w$ and $W\!r$ plotted in Fig.~\ref{writhe} was checked to satisfy
  Eq.~(\ref{Fuller_eq}) with a percentage error $ < 10^{-4}$.  Note also that,
  with the present definition of the centerline as the midpoint of $B_i$ and
  $B_{2n-i}$, a linear DNA chain (closed at infinity) has a finite writhe
  density measured as $W\!r_0/Lk \approx 0.06$. This is due to the corkscrew
  motion of this centerline; a consequence of the fact that the inter-strand
  $B_iB_{2n-i}$ bonds do not cross the center of the tube which tightly
  encloses the equilibrium structure.

\begin{figure*} [ht]
  \includegraphics{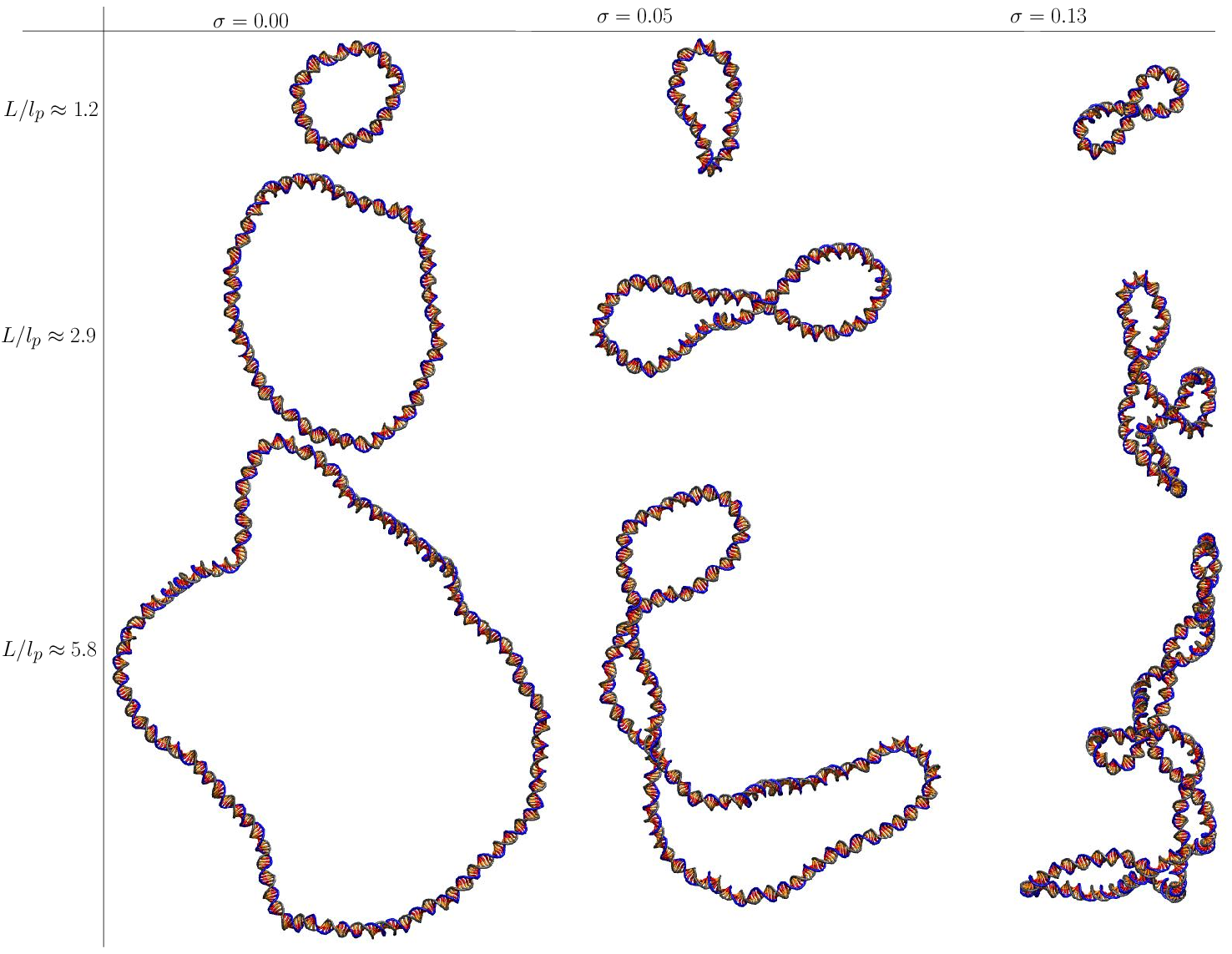}
  \caption{\label{snapshots} Snapshots of model DNA minicircles
    at three different lengths and three different excess linking numbers
    densities ($\sigma$). Note that some of the bonds are omitted in the
     figure for visual clarity.}
\end{figure*}

\subsubsection{Twist and writhe under torsional stress}
On circular DNA chains (such as plasmids) the linking number is a
topological invariant, unaltered by thermal fluctuations ($in\ vivo$,
topoisomers are employed for this reason). The free energy of a
circular DNA of length $L$ is minimized when $Lk = Lk_0 \simeq
L/\lambda$, where, in a strict sense, the equality is attained in the
limit $L\to\infty$ and $T\to 0$. A weak dependence of $\lambda$ on $L$
is possible, but will be ignored here and $\lambda$ is set to the
value we found for our linear model DNA.

In this section, we analyze the behavior of DNA minicircles under
varying linking number. By Eq.~(\ref{Fuller_eq}), an excess or a
deficiency in the linking number ($\Delta Lk = Lk-Lk_0 \neq 0$)
modifies the equilibrium values of writhe and twist measured in the
relaxed state. The partitioning of $\Delta Lk$ among writhe and twist
depends on the stiffnesses associated with the two quantities (i.e.,
bending and twisting rigidities) which are complex functions of the
intra- and inter-strand interactions between the superatoms. Let
\begin{eqnarray} 
  \sigma=\frac{\Delta Lk}{Lk_0} = \frac{Lk} {Lk_0} - 1
\end{eqnarray}
where positive and negative values of $\sigma$ correspond to
overtwisted and undertwisted circular DNA chains, respectively. Note
that, chains of different size are under similar local torsional
stress if their $\sigma$ values are identical. Nevertheless, we show
below that the response of the DNA to $\sigma\neq 0$ depends strongly
on the length.

As discussed above, the circular model DNA chain carries a finite
writhe ($W\!r_0$) and twist ($T\!w_0$) in the relaxed state ($Lk =
Lk_0$) due to the corkscrew motion of the centerline. Since we are
interested in measuring the change in twist and writhe as a function
of $\sigma$, we define the average deviation in twist as
\begin{eqnarray}
  \frac{\langle\Delta T\!w\rangle} {Lk_0} = \frac {\langle T\!w\rangle
    -T\!w_0} {Lk_0}
\end{eqnarray}
and the average deviation in writhe as
\begin{eqnarray}
 \frac{\langle\Delta W\!r\rangle}{Lk_0}  =  \frac {\langle W\!r\rangle -W\!r_0} {Lk_0}
\end{eqnarray}
where the averages are taken over equilibrium snapshots of the system. 

In Fig.~\ref{writhe}, $\langle \Delta Wr \rangle / Lk_0$ and $ \langle
\Delta Tw \rangle / {Lk}_0$ are given as a function of $\sigma$. Four
different chain lengths are considered, $L/l_p\approx$ 1.2, 2.9, 5.8, and
$11.9$ and the results are shown in separate graphs from top-left to
bottom-right, respectively.

Since $Lk$ is an integer, the values of $\sigma$ realizable for a fixed
chain length $L$ are discrete. In order to overcome this constraint, we
combined data obtained from chains with lengths varying upto $\pm 5\%$
of the chosen $L/l_p$ while sampling $\sigma \in [-0.20,0.20]$ by
changing $Lk$. For example, $L/l_p\approx 1.2$ regime was sampled with
$\{Lk = 7,8,\dots,12\} \otimes \{L=105,106,\dots,117\}$. Above
variability in length does not give rise to a significant error in the
average writhe and twist of long chains. For short chains the effect
is more pronounced and fluctuations are observed in the data.
Nevertheless, a general trend that varies with the chain length is
clear in Fig.~\ref{writhe} and will be discussed next.

The average writhe and twist of the circular chains display a
nonmonotonic dependence on $\sigma$ for all four cases considered. Let
us first focus on the shortest chain regime ($L/l_p\approx 1.2$) shown
in the top-left graph of Fig.~\ref{writhe} and the overtwisting scenario with $\sigma > 0$.
There exist two qualitatively distinct modes of torsional response
which are separated by a sharp transition at $\sigma_c^+ \approx
0.07$. For small deviations from $Lk_0$ with $\sigma \! <\!
\sigma_c^+$, the extra linking number is completely absorbed by
the change in twist. In this window the minicircle essentially remains
planar (Fig.~\ref{snapshots}, top-left), with a slightly negative
slope in writhe which is probably associated with the aforementioned
nonzero writhe density of the relaxed chain.

At the transition, the twist drops sharply and the writhe increases
accordingly, absorbing both the additional linking number and the
reduction in twist. The jump in writhe is manifested as an
out-of-plane deformation of the DNA minicircle (Fig.~\ref{snapshots},
top-middle). This phenomenon corresponds to the supercoiling
transition of the minicircle which can be understood through the
buckling instability of the planar state in a circular elastic ribbon
model. In fact, the present data can be used to extract the twist
persistence length ($\equiv K$), of our model DNA through the
relation~\cite{Guitter1992}
\begin{equation}
\sigma_c = \frac{\sqrt{3}\, l_p}{K\cdot Lk_0}
\end{equation}
for an elastic ribbon with a symmetric torsional
response. Substituting $\sigma_c = 0.07$ yields $K/l_p \approx 2.5$ ,
consistent with earlier analysis~\cite{Moroz1997} of the
single-molecule experiments.~\cite{Strick2000} Further increase in
writhe leads to a complete figure-eight shape in the DNA chain as seen
in Fig.~\ref{snapshots} top-right. Beyond the buckling point, the
linking number is almost equally shared by writhe and twist.

A qualitatively similar behavior is observed upon undertwisting in the
interval $0>\sigma>-0.1$.  The buckling transition upon undertwisting
takes place at $\sigma_c^- \approx -0.09$. The asymmetry $|\sigma_c^+|
< |\sigma_c^-|$ suggests that undertwisting is easier than ovetwisting
the DNA chain. Such nonlinear response in torsional stiffness has
already been reported in experiments~\cite{Selvin1992,Bryant2003} and
full-atomistic computer simulations{~\cite{Kannan2006}.  A recent
  analytical model for DNA minicircles by Liverpool {\it et al.}
  ~\cite{Liverpool2008} also predicts that DNA minicircles favor
  supercoiling to denaturation in the weakly nonlinear regime,
  consistent with Fig.~\ref{writhe}. A possible origin of the
  nonlinearity is the presence of twist-bend coupling in chiral
  molecules where the symmetry under 180$^o$ rotation around the helix
  axis is broken (i.e., molecules with major and minor grooves) as
  argued by Marko\&Siggia.~\cite{Marko1994a} Note that, we do not see
  local base-pair hydrogen-bond breaking events~\cite{Harris2008} in
  this regime.  Since our training data set does not include any
  information regarding the melting of the double helical structure,
  it is not surprising that the coarse grained model also does not
  display melting.  Incorporation of full-atomistic single-strand DNA
  simulation data into the model could probably suffice to capture
  such behavior, which is planned as future work.

When we compare the $L/l_p\approx 1.2$ case to longer chains, we
observe two significant differences. First, the buckling transition
takes place at smaller $|\sigma_c|$ values, consistent with
~\citet{Guitter1992}, and practically disappearing for $L/l_p \gtrsim
6.0$. At constant torsional stress density (e.g., $\sigma=0.05$),
minicircles with increasing length accommodate a higher fraction of
$\Delta Lk$ in writhe, with typical configurations shown in the middle
column of Fig.~\ref{snapshots}.

Second, for $L/l_p\approx$ $2.9$, $5.8$, and $11.9$ beyond the
buckling point, the majority of the excess linking number is absorbed
by the writhe, unlike the equal distribution we have seen for
$L/l_p\approx 1.2$. For these longer chains, $\langle \Delta
W\!r\rangle/\langle\Delta T\!w\rangle$ decreases with increasing $|\sigma|$, which
is another manifestation of the nonlinear twist rigidity. For fixed
$\sigma$, $\langle \Delta W\!r\rangle/\langle\Delta T\!w\rangle$ increases
monotonically with length, possibly approaching a finite asymptotic
value.

\subsection{\label{conclusions}Summary and Conclusions}

We presented a coarse-grained model which is designed for studying the
equilibrium structural properties of $10^2$-$10^3$ bps long DNA
minicircles. Proper thermodynamic averaging of global structural
features requires microsecond simulations. Therefore we chose a
minimal, two-bead representation of the sugar-phosphate and the base
in the basic repeat unit.  This approach may be used to address
single-strand chirality and base-flipping which are not accessible to
single-bead models. The model parameters were extracted from
full-atomistic molecular dynamics simulations of DNA oligomers via
Boltzmann inversion. Even at this level of simplicity, used
coarse-graining methodology yields a faithful representation of the
directionality, helicity, major and minor grooves and similar local
characteristic features of DNA. For the sake of simplicity and
efficiency, base-pair specificity and explicit electrostatic
interaction have been ignored here, althought it is straightforward to
incorporate these into the model within the present
approach. Denaturation upon underwinding should be observable after
adding base specificity and including full-atomistic simulations of
single-strand DNA in the Boltzmann inversion step, which will be
considered in a future extension.

Using our model, we performed a systematic molecular dynamics study of
supercoil formation in DNA minicircles. In particular, we measured the
twist/writhe partitioning expressed in Eq.~(\ref{Fuller_eq}) as a function
of the chain length ($L$) and excess linking number density
($\sigma$).  We observed a supercoiling (buckling) transition
associated with the off-plane deformation instability of the
minicircles for $|\sigma|>|\sigma_c^\pm(L)|$, as predicted by
analytical calculations on a simple elastic
model~\cite{Guitter1992,Liverpool2008} and recent full-atomistic
simulations.~\cite{Harris2008} The transition is marked by a sudden
increase in the writhe, while $|\sigma_c|$ decreases with $L$ and
practically disappears beyond $L/l_p \approx 6$ ($\sigma_c^\pm \sim
L^{-1}$ for elastic models). In the planar regime with
$|\sigma|<|\sigma_c^\pm|$, the excess linking number is essentially
stored in the twist.

Our results suggest that, beyond the supercoiling transition, the
fraction of the linking number absorbed as twist and writhe is also
nontrivially dependent on chain length. Chains of the order of a
persistence length carry approximately equal amounts of twist and
writhe, while longer chains accommodate an increasing fraction of the
excess linking number as writhe. The dependence of $\langle\Delta
T\!w\rangle/\langle\Delta W\!r\rangle$ ratio on $\sigma$ remains
nonlinear also for larger chains. At fixed $\sigma$, we observe that
the above ratio increases with $L$ and possibly reaches a finite
asymptotic value in the limit $L\to\infty$, as this limit is
accurately represented by harmonic elastic energy terms for both twist and
writhe. On the other hand, the behavior at fixed $\sigma/L$, which may
apply to the bound portions of a circular DNA chain near the melting
temperature, is unclear and appears to be an interesting problem.

\acknowledgments We gratefully acknowledge funding through TUBITAK Grant
No. 108T553 and Max-Planck-Society Partnership Program for the partner group
with Prof. Kurt Kremer from MPIP-Mainz. A.K. is in debt with D. Mukamel,
E. Orlandini and D. Marenduzzo for helpful discussions and some of the
references.

\end{document}